\documentstyle[prb,aps,preprint]{revtex}
\begin{document}
\draft 
\title{Obtaining Wannier Functions of a Crystalline Insulator within a 
Hartree-Fock approach: applications to LiF and LiCl}
\author{Alok Shukla\cite{email}, Michael Dolg, Peter Fulde} 
\address{Max-Planck-Institut f\"ur
Physik komplexer Systeme,     N\"othnitzer Stra{\ss}e 38 
D-01187 Dresden, Germany}
\author{Hermann Stoll} \address{Institut f\"ur Theoretische Chemie,
Universit\"at Stuttgart, D-70550 Stuttgart, Germany}

\maketitle

\begin{abstract}

An {\em ab initio} Hartree-Fock approach aimed
at directly obtaining the localized orthogonal orbitals (Wannier functions) 
of a 
crystalline insulator is described in detail. The method is used 
to perform all-electron calculations on the ground states of crystalline
lithium fluoride and lithium chloride, without the use of any pseudo or model
potentials. 
Quantities such as total energy, x-ray structure factors and Compton profiles 
obtained using
the localized Hartree-Fock orbitals are shown to be in excellent agreement with
the corresponding quantities calculated using the conventional
Bloch-orbital based Hartree-Fock approach. Localization characteristics of
these orbitals are also discussed in detail.
\end{abstract}
\pacs{ }
\section{INTRODUCTION}
\label{intro}
Electronic-structure calculations on periodic systems are conventionally
done using the so-called Bloch orbital based approach which consists of
assuming an itinerant form for the single-electron wave functions. This
approach has the merit of incorporating the translational invariance of
the system under consideration, as well as its infinite character, in an 
elegant and transparent manner. An alternative approach to
 electronic-structure calculations on periodic systems was proposed by 
Wannier~\cite{wannier}. 
In this approach, instead of describing the electrons in terms of itinerant 
Bloch orbitals, one describes them in terms of mutually orthogonal orbitals
localized on individual atoms or bonds constituting the infinite
solid. Since then such orbitals have come to be known as Wannier functions. 
It can be shown that the two approaches of description of an infinite solid
are completely equivalent and that the two types of orbitals are related by
a unitary transformation~\cite{kohn}. Therefore, the two approaches differ
only in terms of their practical implementation. 
However, the description of metallic systems in terms of Wannier functions
frequently runs into problems as it is found that for such systems the 
decay of the orbitals away from the individual
atomic sites is of power law type and not of exponential type. In other
words, the Wannier functions for such systems are not well 
localized~\cite{kohn}. This behavior is to be expected on intuitive
grounds as electrons in metals are indeed quite delocalized.
On the other hand, for the situations involving surfaces,
impurity states, semiconductors and insulators, where the atomic character of 
electrons is of importance, Wannier functions offer a natural description. 

Recent years have seen an increased amount of activity in the area of
solid-state calculations based on localized orbitals~\cite{wannier2}, of
which Wannier functions are a subclass. 
Most of these approaches have been proposed with the aim of developing 
efficient order-N methods for electronic structure calculations on solids
within the framework of density functional theory. With a different focus,
Nunes and Vanderbilt~\cite{nunes} have developed an entirely
Wannier-function based approach to electronic-structure calculations on
solids in the presence of electric fields, a case for which the eigenstates
of the Hamiltonian are no longer Bloch states. However, we believe that
there is one potential area of application for Wannier orbitals which remains
largely unexplored, namely in the {\em ab initio} treatment of
 electron-correlation effects in solids using the conventional quantum-chemical
methods~\cite{fulde}. It is intuitively obvious that an
{\em ab initio} treatment of electron correlations on large systems will
converge much faster with localized orbitals as compared to delocalized
orbitals because the Coulomb repulsion between two electrons will decay 
rapidly with the increasing distance between the electrons. 
In the quantum-chemistry community the importance of localized orbitals
in treating the correlation effects in large systems was recognized
early on and various procedures aimed at obtaining localized orbitals
were developed~\cite{local}. Some of the localized-orbital approaches were
also carried over to solids chiefly by Kunz and collaborators~\cite{kunz}
at the Hartree-Fock level. This
approach has been applied to a variety of systems~\cite{kunz-r}.
Kunz, Meng and Vail~\cite{kunz-c} have gone beyond the Hartree-Fock
level and also included 
the influence of electron correlations for solids using many-body 
perturbation theory. The scheme of Kunz et al. 
is based upon nonorthogonal orbitals which, in general, are better localized
than their orthogonal counterparts. However, the
subsequent treatment of electron correlations with nonorthogonal orbitals 
is generally much more complicated than the one based upon true Wannier 
functions.

 In our group electron correlation effects on solids have been studied
using the incremental scheme of Stoll~\cite{increment}
which works with localized orbitals. In such studies
the infinite solid is modeled as a large enough cluster and then correlation 
effects are calculated by incrementally correlating the Hartree-Fock reference
state of the cluster expressed in terms of localized orbitals~\cite{corr}.
However, a possible drawback of this procedure is that there will always
be finite size effects and no {\em a priori} knowledge is available 
as to the difference in results when compared with the infinite-solid limit.
In order to be able to study electron-correlation effects in 
the infinite-solid limit using conventional quantum-chemical approaches, 
one first has to obtain a Hartree-Fock representation of the system 
in terms of Wannier functions. This task is rather complicated because, 
in addition to the localization requirement, one also imposes the 
constraint upon the Wannier functions that
they be obtained by the Hartree-Fock minimization of the total energy of the
infinite solid. In an earlier paper~\cite{shukla}---henceforth referred to
as I---we had outlined precisely such a procedure which obtained the
Wannier functions of an infinite insulator within a  Hartree-Fock approach
and reported its preliminary applications to the lithium
hydride crystal. In the present paper we describe all 
theoretical and computational details of the approach and report
applications to larger systems namely lithium fluoride and lithium
chloride. Unlike I, where we only reported results on the total energy
per unit cell of the system, here we also use the Hartree-Fock Wannier 
functions to compute the x-ray structure 
factors and Compton profiles. Additionally, we also discuss the localization
characteristics of the Wannier functions in detail. All the physical
quantities computed with our procedure are found to be in excellent agreement
with those computed using the CRYSTAL program~\cite{crystalprog} which 
employs a Bloch orbital based {\em ab initio} Hartree-Fock approach.
In a future publication we will apply the present formalism to 
perform {\em ab initio} correlation calculations on an infinite insulator.

The rest of this paper is organized as follows. In section \ref{theory} we
develop the theoretical formalism at the Hartree-Fock level by minimizing
the corresponding energy functional, coupled with the 
requirement of translational symmetry, and demonstrate that the resulting HF equations correspond
to the HF equations for a unit cell of the solid
embedded in the field of identical unit cells constituting the rest of the 
infinite solid. Thus an embedded-cluster picture for the infinite 
solid emerges rigorously from this
derivation.  Subsequently a localizing potential is introduced in the 
HF equations by means of projection operators leading to our working 
equations for the Hartree-Fock Wannier orbitals for an infinite solid.
Finally, these equations are cast in the matrix form using a linear combination
of atomic orbitals approach which is used in the actual calculations. 
In section \ref{results} we present the results of our calculations
performed using the aforementioned formalism on LiF and 
LiCl crystals. 
Finally, in section \ref{conclusion} we present our conclusions. Various 
aspects related to the computer implementation of the present approach are 
discussed in the appendix.
\section{THEORY}
\label{theory}
\subsection{Hartree-Fock Equations}
\label{hft}
We consider the case of a perfect solid without the presence of any
impurities or lattice deformations such as phonons. We also ignore the
effects of relativity completely so that the spin-orbit coupling is
also excluded. In such a case, in atomic
units~\cite{au}, 
the nonrelativistic Hamiltonian of the system consisting of the kinetic 
energy of electrons, 
electron-nucleus
interaction, electron-electron repulsion and nucleus-nucleus
interaction is given by
\begin{equation}
H = - \frac{1}{2}\sum_{i} \nabla_{i}^{2} - \sum_{i} \sum_{I}
\frac{Z_{I} }{|\bf{r}_{i} - \bf{R}_{I}|} + \sum_{i > j} \frac{1}
{|\bf{r}_{i}-\bf{r}_{j}|} + \sum_{I>J}\frac{Z_{I} Z_{J}}
{|\bf{R}_{I} - \bf{R}_{J}|}
 \mbox{,} 
\label{eq-ham}
\end{equation}
where in the equation above $\bf{r}_i$ denotes the position coordinates
of the $i$-th electron while $\bf{R}_{I}$ and $Z_{I}$ respectively denote the 
position and the charge of the $I$-th nucleus of the lattice. For a given
geometry of the solid the last term representing the nucleus-nucleus 
interaction will make a constant contribution to the energy and will not 
affect the dynamics of the electrons. To develop the theory further we
make the assumptions that the solid under consideration is a closed-shell
system and that a single Slater determinant represents a reasonable
 approximation
to its ground state. Moreover, we assume that the same spatial orbitals 
represent both the spin projections of a given shell, i.e., we confine
ourselves to restricted Hartree-Fock (RHF) theory. With the preceding
assumptions, the total energy of the solid can be written
as 
\begin{equation}
E_{solid} = 2 \sum_{i} <i|T|i> +  2 \sum_{i} <i|U|i> + \sum_{i,j} ( 2 <ij|ij>
 - <ij|ji>)+E_{nuc} \mbox{,} 
\label{eq-esolid}
\end{equation}
where $|i>$ and $|j>$ denote the occupied spatial orbitals assumed to form an
orthonormal set, 
$T$ denotes the 
kinetic energy operator, $U$ denotes the electron-nucleus potential 
energy, $E_{nuc}$ denotes the nucleus-nucleus interaction energy and $<ij|ij>$ etc. represent the two-electron integrals involving the
electron repulsion. 
The equation above is completely independent of
the spin degree of freedom which, in the absence of spin-orbit coupling, can
be summed away leading to familiar
factors of two in front of different terms.
Clearly the terms involving $<i|U|i>$, $<ij|ij>$, and $E_{nuc}$
contain infinite lattice sums and are convergent only when combined together.
So far the energy expression of Eq.(\ref{eq-esolid}) does not incorporate any
assumptions regarding the translational symmetry of a perfect solid. 
In keeping with our desire to introduce translational symmetry in the
real space, without having to invoke the {\bf k}-space as is usually done in 
the Bloch orbital based theories, we make the following observation.
A crystalline solid, in its ground state, is composed of identical unit cells 
and the orbitals belonging to a given unit cell are identical to the 
corresponding orbitals belonging to any other unit cell and are related to 
one another by a simple translation operation. Assuming that the number of
orbitals in a unit cell is $n_{c}$ and if we denote the $\alpha$-th orbital 
of a unit cell located at the position given by the vector ${\bf R}_{j}$ of 
the lattice by $|\alpha({\bf R}_{j})>$ then clearly the   
set $\{ |\alpha({\bf R}_{j})>; \alpha =1,n_{c}; j=1,N \}$  
denotes all the orbitals of the solid.  In the previous expression $N$ is 
the total number of unit cells in the solid which, of course, is infinite.
Henceforth, Greek labels $\alpha, \beta, \gamma, \ldots$ will always denote the
orbitals of a unit cell. 
The translational symmetry condition expressed in the real space can be stated 
simply as
\begin{equation}
|\alpha({\bf R}_{i}+{\bf R}_{j})> = {\cal T} ({\bf R}_{i}) 
|\alpha({\bf R}_{j})> \mbox{,}
\label{eq-trsym}
\end{equation}
where ${\cal T} ({\bf R}_{i})$ is an operator which represents a translation
by vector ${\bf R}_{i}$. Using  this, one can rewrite the energy 
expression of Eq.(\ref{eq-esolid}) as
\begin{eqnarray}
E         & = & N \left \{ 2 \sum_{\alpha=1}^{n_{c}} <\alpha(o)|T|\alpha(o)> +
   2 \sum_{\alpha=1}^{n_{c}} <\alpha(o)|U|\alpha(o)>  \right. \nonumber  \\ 
          &   & + \left.  \sum_{\alpha,\beta=1}^{n_{c}} \sum_{j=1}^{N}
 ( 2 <\alpha(o)\beta({\bf R}_{j})|\alpha(o)\beta({\bf R}_{j})>
 - <\alpha(o)\beta({\bf R}_{j})|\beta({\bf R}_{j})\alpha(o)>) + e_{nuc} 
\right\} 
\mbox{,} 
\label{eq-esolidf}
\end{eqnarray}
where $|\alpha(o)>$ denotes an orbital centered in the reference unit cell,
$e_{nuc}$ involves the interaction energy of the nuclei of the reference
cell with those of the rest of the solid ($E_{nuc}=Ne_{nuc}$), and
we have removed the subscript $solid$ from the energy. The preceding
equation also assumes the important fact that the orbitals obtained
by translation operation of Eq.(\ref{eq-trsym}) are orthogonal to each
other. We shall elaborate this point later in this section. An important 
simplification to be noted here is that by assuming the translational 
invariance
in real space as embodied in Eq.(\ref{eq-trsym}), we have managed to express
the total Hartree-Fock energy of the infinite solid in terms of a finite
number of orbitals, namely the orbitals of a unit cell $n_{c}$. If we require
that the energy of Eq.(\ref{eq-esolidf}) be stationary with 
respect to the first-order variations
%in the orbitals of a unit cell, subject to the orthogonality constraint,
in the orbitals, subject to the orthogonality constraint,
%we obtain the desired Hartree-Fock equations
we are led to the Hartree-Fock operator 
\begin{equation}
% ( T + U
% + 2 \sum_{\beta} J_{\beta}- \sum_{\beta} K_{\beta} ) |\alpha>
% = \epsilon_{\alpha} |\alpha>
 H_{HF}=T + U
 + 2 \sum_{\beta} J_{\beta}- \sum_{\beta} K_{\beta} 
\label{eq-hff}         
\end{equation}  
where J and K---the conventional Coulomb and exchange 
operators, respectively---are defined as
\begin{equation}
\left.
 \begin{array}{lll}
 J_{\beta}|\alpha> & = & \sum_{j} <\beta({\bf R}_{j})|\frac{1}{r_{12}}|
\beta({\bf R}_{j})>|\alpha> \\  
 K_{\beta}|\alpha> & = & \sum_{j} <\beta({\bf R}_{j})|\frac{1}{r_{12}}|\alpha>
|
\beta({\bf R}_{j})> \\  
\end{array}
 \right\}  \mbox{.} \label{eq-jk} 
\end{equation}
Any summation  over Greek indices $\alpha, \beta, \gamma, \ldots$ 
will imply summation
over all the $n_{c}$ orbitals of a unit cell unless otherwise specified. 
%For the sake of brevity,  $|\alpha>$ will denote
%$|\alpha(o)>$, an orbital centered in the reference unit cell. As mentioned
As mentioned
earlier, the terms $U$, $J$ and $K$ involve infinite lattice sums and their
%practical evaluation will be discussed in the next section. Eq.(\ref{eq-hff})
%is the canonical Hartree-Fock equation for the orbitals of the reference unit 
%cell and therefore its solutions will be orthogonal to each other. However,
practical evaluation will be discussed in the next section. 
The eigenvectors of the Hartree-Fock operator of Eq.(\ref{eq-hff}) will be orthogonal to each other, of course. However,
%in general, its solutions would neither be localized, nor would they be 
in general, these solutions would neither be localized, nor would they be 
orthogonal to the orbitals of any other unit cell. 
This is because the orbitals centered in 
any other unit cell are obtained from those of the reference cell
using a simple translation operation as defined in Eq.(\ref{eq-trsym}), which 
does not impose any orthogonality or localization constraint upon them. 
Since our aim is to 
obtain the Wannier functions of the infinite solid, i.e., all the orbitals
of the solid must be localized and orthogonal to each other, we will 
%have to impose  these requirements explicitly upon the solutions 
have to impose  these requirements explicitly upon the eigenspace
%of Eq.(\ref{eq-hff}). 
of (\ref{eq-hff}). 
This can most simply be accomplished by including in  
%Eq.(\ref{eq-hff}) the projection operators
(\ref{eq-hff}) the projection operators
corresponding to the orbitals centered in the unit cells in a
(sufficiently large) neighborhood of the reference cell 
\begin{equation}
( T + U
 + 2  J-  K   
+\sum_{k \in{\cal N}} \sum_{\gamma} \lambda_{\gamma}^{k} 
|\gamma({\bf R}_{k})>
<\gamma({\bf R}_{k})| ) |\alpha>
 = \epsilon_{\alpha} |\alpha>
\mbox{,}
\label{eq-hff1}         
\end{equation}  
%where $J= \sum_{\beta} J_{\beta}$, $K= \sum_{\beta} K_{\beta}$
where $|\alpha>$ stands for
$|\alpha(o)>$, an orbital centered in the reference unit cell,
$J= \sum_{\beta} J_{\beta}$, $K= \sum_{\beta} K_{\beta}$, 
and ${\cal N}$ collectively denotes the unit cells in the aforementioned
neighborhood. Clearly the choice of ${\cal N}$ 
will be dictated by the system under consideration---the more delocalized
 electrons
of the system are, the larger will ${\cal N}$ need to be. 
In our calculations we have typically chosen ${\cal N}$ to include
up to third nearest-neighbor unit cells of the reference cell. 
In the equation above $\lambda_{\gamma}^{k}$'s are the shift parameters
associated with the correponding orbitals of ${\cal N}$. 
For perfect orthogonality and localization, their values should be infinitely
high. By setting the shift parameters $\lambda_{\gamma}^{k}$'s to infinity
we in effect raise the orbitals localized in the environment unit cells
(region ${\cal N}$) to very high energies compared to those localized
in the reference cell. Thus the lowest energy solutions of Eq.(\ref{eq-hff1})
will be the ones which are localized in the reference unit cell and are
orthogonal to the orbitals of the environment cells. 
Of course, in practice, it suffices to choose a rather large value
for these parameters, and the issue pertaining to this numerical choice 
is discussed further in section \ref{results}. 
Eq.(\ref{eq-hff1}) will generally be solved iteratively
as described in the next section. If the initial guesses for the orbitals
of the unit cell $\{|\alpha>, \alpha=1,n_{c}> \}$ are localized, subsequent
orthogonalization by means of projection operators will not destroy that
property~\cite{local} and the final solutions of the problem will be 
localized orthogonal orbitals.
Therefore, projection operators along with the shift parameters,
simply play the role of a localizing potential~\cite{local} as it is clear 
that upon convergence their contribution to the Hartree-Fock equation vanishes.
The orbitals contained in unit cells located farther 
than those in ${\cal N}$ should be  automatically orthogonal to
the reference cell orbitals by virtue of the large distance between them. 
It is clear that the orthogonalization of the orbitals to each other 
will introduce oscillations in these orbitals which are also referred to
as the orthogonalization tails.

Combining the orthogonality of the neighboring orbitals to the reference cell 
orbitals with the translation symmetry of the infinite solid, it is easy
to see that the orbitals of any unit cell are orthogonal 
to all the orbitals of the rest of the unit cells. Therefore, orbitals thus
obtained are essentially Wannier functions.
After solving for the HF equations presented above one can obtain 
the electronic part of the energy per unit cell simply by dividing
the total energy of Eq.(\ref{eq-esolidf}) by $N$, 
which, unlike the total energy, is a finite quantity.

In paper I we arrived at exactly the same HF equations as above, although 
we had followed a more intuitive path utilizing the so-called 
``embedded-cluster'' philosophy,  whereby we minimized only that 
portion of the total
energy of Eq.(\ref{eq-esolid}) which corresponds to the 
``cluster-environment'' interaction. The fact that the derivations reported
in paper I, and here, both lead to the same final equations has to do with
the translation invariance which allows the total energy to be expressed
in the form of Eq.(\ref{eq-esolidf}).
Therefore, we emphasize that the equations derived above are exact   
and do not involve any approximation other than the Hartree-Fock approximation
itself. Thus results of all the computations utilizing this approach should 
be in 
complete agreement with the equivalent computations performed using
the traditional Bloch orbital based approach as is implemented, e.g, in the
program CRYSTAL~\cite{crystalprog}.

 By inspection of Eq.(\ref{eq-hff1}) it is clear
that it is of the embedded-cluster form in the sense that if one
calls the reference unit cell the ``central cluster'', it describes the
dynamics of the electrons of this central cluster embedded in the field of
identical unit cells of its environment (rest of the infinite solid).
\subsection{Linear Combination of Atomic Orbital Implementation}
\label{lcao}
We have performed a computer implementation of the formalism presented 
in the previous section within a linear combination of atomic orbital
(LCAO) approach, whereby we transform the differential equations of
Eq.(\ref{eq-hff1}) into a set of linear equations solvable by matrix
methods. Atomic units were used throughout the numerical work. 
We proceed by expanding the orbitals localized in the reference cell as
\begin{equation}
|\alpha> =  \sum_{p} \sum_{{\bf R}_{j} \in {{\cal C} + \cal N} } 
C_{p,\alpha} |p({\bf R}_{j})>  \: \mbox{,}
\label{eq-lcao}
\end{equation}
where ${\cal C}$ has been used to denote the reference cell, ${\bf R}_{j}$ 
represents the location of the  
$j$th unit cell (located in ${\cal C}$ or ${\cal N}$)  
and $|p({\bf R}_{j})>$ represents
a basis function centred in the $j$th unit cell. In order to account for the
orthogonalization tails of the reference cell orbitals, it is necessary to
include the basis functions centred in ${\cal N}$ as well. Clearly, 
the translational symmetry of 
the crystal as expressed in Eq.(\ref{eq-trsym}) demands that 
the orbitals localized in two different unit cells have
the same expansion coefficients $C_{p,\alpha}$, and differ only in
the location of the centers of the basis functions. The LCAO formalism
implemented in most of the quantum-chemistry molecular programs, as also in 
the CRYSTAL code\cite{crystalprog}, expresses the basis functions
$|p({\bf R}_{j})>$ of Eq.(\ref{eq-lcao}) as linear combinations of 
Cartesian Gaussian type basis functions (CGTFs)  of the 
form 
\begin{equation}
\phi({\bf r}, \eta_{p}, {\bf n}, {\bf R})= (x-R_{x})^{n_x}(y-R_{y})^{n_y} 
                                            (z-R_{z})^{n_z} \mbox{exp}(-\eta_{p}
({\bf r}-{\bf R})^2)
\mbox{,}
\label{eq-cgto}
\end{equation}
where ${\bf n}=(n_{x},n_{y},n_{z})$. In the previous equation, $\eta_{p}$ 
denotes the exponent and the vector 
${\bf R}$ represents the center of the basis 
function. The centers of the basis functions ${\bf R}$ are normally taken to
be at the locations of the appropriate atoms of the system. 
CGTFs with $n_{x}+n_{y}+n_{z}=0,1,2,\ldots$ are called
respectively $s, p, d \ldots$ type basis functions\cite{ref-bas}. The individual
basis functions of the form of Eq.(\ref{eq-cgto}) are called
{\em primitive} functions while the linear combinations of them are
called the {\em contracted} functions. 
The formalism is 
totally independent of the type of basis functions, but for the sake of 
computational simplicity, we have programmed our approach using Gaussian 
lobe-type functions\cite{whitten}. In this approach one approximates
the $p$ and higher angular momentum CGTFs as linear combinations
of $s$-type basis functions displaced by a small amount from the
location of the atom concerned. For example, in the present study 
a primitive $p$ type CGTF 
centered at the origin was approximated as
\begin{equation}
\phi_{p}({\bf r}, \eta)  =  A\{\mbox{exp}(-\eta({\bf r}+{\bf d}_{\eta})^{2})
 - \mbox{exp}(-\eta({\bf r}-{\bf d}_{\eta})^{2})\}
\mbox{,}
\label{eq-lobes}
\end{equation}
where, $A$ is the normalization constant and $|{\bf d}_{\eta}|= C/ 
\sqrt{\eta}$. In the present study the value of 0.1 atomic units (a.u.) was 
employed for $C$. For approximating the $p_{x}$, $p_{y}$ and $p_{z}$ types of 
basis functions, the displacement vectors ${\bf d}_{\eta}$ are chosen to be
along the positive $x$, $y$ and $z$ directions, respectively.
By substituting Eq.(\ref{eq-lcao}) in Eq.(\ref{eq-hff1}) we obtain the HF 
equations in the LCAO matrix form   
\begin{equation}
\sum_{q} F_{pq} C_{q,a} =  \epsilon_{a} \sum_{q} S_{pq} C_{q,a}
\mbox{.}
\label{eq-scf}
\end{equation} 
The Fock matrix $F_{pq}$ occuring in the equation above is defined as 
\begin{equation}
F_{pq}  = 
<p | ( T + U
 + 2 J - K) |q> + \sum_{k \in{\cal N}} \sum_{\gamma}  
\sum_{p',q'} \lambda_{\gamma}^{k} S_{pp'}S_{qq'}C_{p',\gamma}
C_{q',\gamma}  
\mbox{,}
\label{eq-fock}
\end{equation}
where the contribution of all the operators appearing in Eq.(\ref{eq-hff1})
has been replaced by the corresponding matrices in the representation
of the chosen basis set. Above, unprimed functions $|p>$ and $|q>$  represent the basis functions 
corresponding to the orbitals of the reference unit cell while the primed 
functions $|p'>$ and $|q'>$ denote the basis functions corresponding to the 
orbitals of ${\cal N}$. 
In particular, the overlap matrix is given by
\begin{equation}
S_{pq} = <p|q>
\label{eq-ovp}
\end{equation} 
and the Coulomb and the exchange matrix elements are defined as
\begin{equation}
<p|J |q>= \sum_{r,s} \sum_{k} <p \: r({\bf R}_{k})|\frac{1}{r_{12}}|q \:
s({\bf R}_{k})> D_{rs} 
\label{eq-jpq}
\end{equation}
and 
\begin{equation}
<p|K |q>= \sum_{r,s} \sum_{k} <p \: r({\bf R}_{k})|\frac{1}{r_{12}}|s({\bf
 R}_{k}) \: q > D_{rs} 
\mbox{,}
\label{eq-kpq}
\end{equation}
where $D_{rs}$ denotes the elements of the density matrix $D$ of the orbitals
of a unit cell evaluated as~\cite{denmat}
\begin{equation}
D_{pq}= \sum_{\alpha} C_{p,\alpha} C_{q,\alpha}
\mbox{.}
\label{eq-denmat}
\end{equation}
The matrix form of the HF equations (\ref{eq-scf}) is  a pseudo 
eigenvalue problem which can be solved iteratively to obtain the HF orbitals. 
The energy per unit cell can be computed by means of a simple matrix-trace
 operation 
\begin{eqnarray}
E_{cell} & = &  Tr\{(2 T +  2 U 
 + 2 J  - K)D\}    
+ e_{nuc}    
 \mbox{,} 
\label{eq-ecell}
\end{eqnarray}
where above $T$, $U$, $J$ and $K$ and $D$ denote the matrices of the
corresponding operators in the representation of the chosen basis set, and 
$e_{nuc}$ was defined after Eq.(\ref{eq-esolidf}).

In practice one proceeds according to the following algorithm:
\begin{enumerate}
\item Start with some localized initial guess for the orbitals of the 
reference cell. For ionic systems considered here we chose these to be the 
orbitals of the individual ions centered on the corresponding atomic
sites. For covalent systems, it would be reasonable to use
suitable bonding combinations of atomic orbitals.
\item Use these orbitals to construct the Fock matrix as defined in 
Eq.(\ref{eq-fock}).
\item Diagonalize the Fock matrix to obtain a new set of orbitals of  
     the reference cell. 
\item Compute the energy per unit cell by using Eq.(\ref{eq-ecell}).
\item Go to step 2. Iterate until the energy per unit cell has converged. 
\end{enumerate} 
Various mathematical formulas and computational aspects related to the
evaluation of different contributions to the Fock matrix are discussed in the
appendix.
\subsection{Evaluation of Properties}
\label{sec-prop}
In this section we describe the evaluation of the x-ray structure factors
and Compton profiles from the Hartree-Fock Wannier functions obtained
from the formalism of the previous section. Both these properties
can be obtained from the first-order density matrix of the system
defined for the present case as
\begin{equation}
\rho({\bf r},{\bf r}') = 2 \sum_{i} \sum_{\alpha} \phi_{\alpha}({\bf r}-  
{\bf R}_{i}) \phi_{\alpha}^{*}({\bf r}'-  {\bf R}_{i})
\label{eq-den}
\end{equation}
where $\phi_{\alpha}({\bf r}-  {\bf R}_{i}) = <{\bf r}|\alpha({\bf R}_{i})>$ is the 
$\alpha$-th HF orbital of the unit cell located at position ${\bf R}_{i}$.
The factor of two above is a consequence of spin summation.
\subsubsection{X-ray Structure Factors}
\label{sec-xfac}
By measuring the x-ray structure factors experimentally one can obtain useful
information on the
charge density of the constituent electrons. Theoretically, the x-ray structure factor $S({\bf k})$ can be obtained by taking the Fourier
 transform of the diagonal part of the first-order density matrix
\begin{equation}
S({\bf k})= \int \rho({\bf r},{\bf r})\mbox{exp}(i{\bf k}\cdot {\bf r}) d{\bf
  r}
\label{eq-xfac} 
\end{equation}
\subsubsection{Compton Profile}
\label{sec-cprof}
By means of Compton scattering based experiments, one can extract the 
information on the momentum distribution of the electrons of the solid.
In the present study we compute the Compton profile in the impulse 
approximation as developed by Eisenberger and Platzman~\cite{cp-imp}. Under
the impulse approximation the Compton profile for the momentum transfer $q$
is defined as~\cite{cp-imp}
\begin{equation}
J(q)=\frac{1}{(2\pi)^{3}} \int \delta(\omega - \frac{k^2}{2m} -\frac{q}{m})
M({\bf p}) d {\bf p}
\mbox{,}
\label{eq-cp1}
\end{equation}
where ${\bf k}$ and $\omega$ are, respectively, the changes in momentum and 
the frequency of the incoming x- or $\gamma$-ray due to scattering,   
${\bf p}$ is the Compton electron momentum, $q=\frac{{\bf k}\cdot {\bf
p}}{k}$ is the projection of ${\bf p}$ in the direction of ${\bf k}$, the
delta function imposes the energy conservation  and
$M({\bf p}) = \rho({\bf p},{\bf p})$ denotes the electron momentum distribution obtained
from the diagonal part of the full Fourier transform of the first-order 
density matrix  
\begin{equation}
\rho({\bf p},{\bf p}') = \int \mbox{exp}
(i({\bf p}\cdot{\bf r} - {\bf p}'\cdot{\bf r}')) \rho({\bf r},{\bf r}')d{\bf r}
d{\bf r}'
\mbox{.}
\label{eq-emd}
\end{equation}
By choosing the $z$-axis of the coordinate system defining ${\bf p}$ along the
direction of ${\bf k}$, one can perform the $p_{z}$ integral in
Eq.(\ref{eq-cp1}) to yield
\begin{equation}
J(q)=\frac{1}{(2\pi)^{3}} \int_{-\infty}^{\infty} dp_{x} 
\int_{-\infty}^{\infty} dp_{y} 
M(p_{x},p_{y},q) 
\mbox{,}
\label{eq-cp}
\end{equation}

 Integrals contained in the expressions for the x-ray structure factor and the
Compton profile (Eqs. (\ref{eq-xfac}) and (\ref{eq-cp}), respectively) can be performed
analytically when the density matrix is represented in terms of Gaussian
lobe-type basis functions. These analytic expressions are used to evaluate the
quantities of interest in our computer code, once the Hartree-Fock density
matrix has been determined.
\section{CALCULATIONS AND RESULTS}
\label{results}
In this section we present the results of the calculations performed on
crystalline LiF and LiCl. Prencipe 
et al.~\cite{prencipe} studied these compounds, along with  
several other alkali halides, using the CRYSTAL program~\cite{crystalprog}.
CRYSTAL,  as mentioned earlier, is a Bloch orbital 
based {\em ab initio} Hartree-Fock program set up within an LCAO scheme, 
utilizing CGTFs as basis functions. In their study, Prencipe et al.
employed a very large basis set and, therefore, their results are believed to 
be very close to the Hartree-Fock limit. In the present work
our intention is not to repeat the extensive calculations of 
Prencipe et al.~\cite{prencipe}, but rather to demonstrate  
that at the Hartree-Fock level one can obtain the 
same physical insights by applying  the Wannier function based approach as 
one would by utilizing the  Bloch orbital based approach. Moreover,
because of the use of lobe functions as basis functions, we run into
problems related to numerical instability when very diffuse $p-$type 
(and beyond) basis functions
are employed. In future we intend to incorporate true CGTFs as basis functions
in our program, which should make the code  numerically much more stable.
Therefore, we have performed 
these calculations with modest sized basis sets. We reserve the use of
large basis sets for the future calculations, when we intend to go beyond the 
Hartree-Fock level to utilize these Wannier functions to do correlated
calculations.  The reason we have chosen to compare our results to those
obtained using the CRYSTAL program is because not only is CRYSTAL based upon
an LCAO formalism employing Gaussian type of basis functions similar
to our case, but
also it is a well-tested program and widely believed to be the state of 
the art in crystalline Hartree-Fock calculations~\cite{crystalbook}. 

All the calculations to be presented below assume the observed  
face-centered cubic (fcc) structure for the compounds. 
The reference unit cell ${\cal C}$ was taken to be the primitive cell
containing an anion at the $(0,0,0)$ position and the cation at 
$(0,0,a/2)$, where $a$ is the lattice constant. 
The calculations were performed with different values of the lattice constants
 to be indicated later. The basis sets used for lithium, fluorine and chlorine
are shown in tables \ref{tab-basli}, \ref{tab-basf} and \ref{tab-bascl}, 
respectively. 
For lithium we adopted the basis set of Dovesi et al. used in their
lithium hydride study~\cite{lih-dovesi}, while for fluorine and chlorine
basis sets originally published by Huzinaga and collaborators~\cite{huzbas} 
were used.
The values of the level-shift parameters $\lambda_{\gamma}^{k}$'s of 
Eq.(\ref{eq-fock})
should be high enough to guarantee sufficient orthogonality while still
allowing for numerical stability. Thus this choice leaves sufficient room
for experimentation. We found the values in the range 
$\approx 1.0  \times 10^{3}\mbox{---} 1.0 \times 10^{4} \; \mbox{a.u.}$ suitable for our 
work. We verified by explicit calculations that our results had indeed
converged with respect to the values of the shift parameters. In the course of the evaluation of integrals needed to construct
the Fock matrix, all the integrals whose magnitudes were smaller than 
$1.0  \times 10^{-7} \; \mbox{a.u.}$ were discarded both in our calculations
as well as in the CRYSTAL calculations.

The comparison of our ground-state energies per unit cell with those obtained
using the identical basis sets by the CRYSTAL program~\cite{crystalprog}
is illustrated in tables \ref{tab-enlif} and \ref{tab-enlicl} for different
values of lattice constants. The biggest disagreement
between the two types of calculations is 0.7 
millihartree. A possible source of this disagreement is our use of lobe
functions to approximate the $p$-type CGTFs. However,  
since the typical accuracy of a CRYSTAL calculation is also
1 millihartree~\cite{crystalprog}, we consider this disagreement to 
be insignificant. Such excellent agreement between the total energies 
obtained using two different approaches gives us confidence as to the
essential correctness of our approach. From the results it is also
obvious that the basis set used in these calculations is inadequate to
predict the lattice constant and the bulk modulus correctly. To be
able to do so accurately, one will have to employ a much larger basis set such
as the one used by Prencipe et al.~\cite{prencipe}. Since Hartree-Fock
lattice constants generally are much larger than the experimental value,
 we reserve the large-scale Hartree-Fock calculations for
future studies in which we will also go beyond the Hartree-Fock level to 
include the influence of electron correlations.

 Valence Wannier functions for LiF and LiCl  are plotted along different
crystal directions in Figs.\ \ref{fig-lif2p-001}, 
\ref{fig-lif2p-110}, \ref{fig-licl3p-001} and \ref{fig-licl3p-110}. 
Lattice constants for these calculations were assigned their experimental
values~\cite{wyckoff} of 3.99 $\AA$  and 5.07 $\AA$, for LiF and LiCl,
respectively. Although core orbitals were also obtained from the same  
set of calculations,  we have not plotted them here because they are 
trivially localized. The $p$-character of the Wannier functions is evident
from the antisymmetric nature  of the plots under reflection. The additional 
nodes introduced in the orbitals due to their orthogonalization to orbitals 
centered on the atoms of region ${\cal N}$ are also evident. The localized 
nature of these orbitals is obvious from the fact that the orbitals
decay rapidly as one moves away from the atom under consideration. The
orthogonality of the orbitals of the reference cell to those of the
neighborhood (region ${\cal N}$) was always better than $1.0 \times 10^{-5}$.

Now we discuss the data for x-ray structure factors. These quantities were
also evaluated at experimental lattice constants mentioned above.
The x-ray structure factors obtained by our method are compared to values
calculated with the CRYSTAL program, and experimental data, 
in tables \ref{tab-lifxfac} 
and \ref{tab-liclxfac} for LiF and LiCl, respectively. For the case of LiF we
directly compare the theoretical values with the experimental data of 
Merisalo et al.~\cite{exp-xraylif}, 
extrapolated to zero temperature by Euwema et al.~\cite{exp-xraylifeuw}.
For LiCl it was not possible for us to extrapolate the experimental data of 
Inkinen et al.~\cite{exp-xraylicl}, measured at
 $T= 78\; K$, to the corresponding zero temperature values. Therefore,
to compare our LiCl calculations to the experiment,
we correct our theoretical values for thermal motion using the
Debye-Waller factors of $B_{\mbox{Li}}=0.93 \AA^{2}$ and $B_{\mbox{Cl}}=0.41 
\AA^{2}$, measured also by Inkinen et al.~\cite{exp-xraylicl} The Debye-Waller
corrections were applied to the individual form factors of Li$^{+}$ and
Cl$^{-}$ ions. 
 From both the tables it is obvious that our results
are in almost exact agreement with those of CRYSTAL. This implies that
our Wannier function HF approach based description of the charge density of 
systems considered here, is identical to a Bloch orbital based HF description 
as formulated in CRYSTAL~\cite{crystalprog}. For the case of LiF the 
agreement between our results and the experiment is also quite good, maximum
errors being $\approx$ 5\%. For the case of LiCl, our corrected values
of x-ray structure factors deviate from the experimental values at  most 
by approximately 3\%. 
Perhaps by using a larger basis set one can obtain even better agreement 
with experiments. 

Finally we turn to the discussion of Compton profiles. 
Directional and isotropic Compton profiles, computed using our 
approach and the CRYSTAL program, are compared to the isotropic Compton profiles
measured by Paakkari et al.~\cite{exp-cp},  in tables \ref{tab-lifcp} and 
\ref{tab-liclcp}. We obtain the isotropic Compton profiles from our 
directional 
profiles by performing a directional average of the profiles
along the three crystal
directions according to the formula $<J> = \frac{1}{26}(6J_{100}+12 J_{110}
+ 8 J_{111})$ valid for an fcc lattice~\cite{exp-dcp}.  
While experimental data for 
directional Compton profiles exist in the case of LiF~\cite{exp-dcp},
no such measurements have been performed for LiCl, to the best of our knowledge.
For LiF there is close agreement between our results and the
ones calculated using the CRYSTAL program. For LiCl our results disagree with
the CRYSTAL results somewhat for small values of momentum transfer, although 
relatively 
speaking the disagreement
is quite small---the maximum deviation being $\approx$ 0.3\% for $q=0.0$ and
the $[100]$ direction. The possible source of the disagreement may be that to
get the values of Compton profiles for all the desired values of
momentum transfer, we had
to use the option of CRYSTAL~\cite{crystalprog} where the Compton profiles
are obtained by using the real-space density matrix rather than its more
accurate {\bf k}-space counterpart. However, as is clear from the tables,
even for those worst cases, there is no significant difference between
the averaged out isotropic Compton profiles obtained in our computations
and those obtained from CRYSTAL. At the larger values of momentum transfer,
our results are virtually identical to the CRYSTAL results. The close
agreement with CRYSTAL clearly implies that our Wannier function based
description of the momentum distribution of the electrons in the solid
is identical
to the one based upon Bloch orbitals.

 Considering the fact that we have used a rather modest basis set,
it is quite surprising that the values of isotropic Compton profiles 
obtained by us are in close
agreement with the corresponding experimental values~\cite{exp-cp}. An 
inspection of tables \ref{tab-lifcp} and 
\ref{tab-liclcp} reveals that the calculated values always agree with the 
experimental ones to within 6\%. However, ours as well as the CRYSTAL
 calculations presented here are not able to describe the observed anisotropies
in the directional Compton profiles~\cite{exp-dcp} for LiF which is also the
reason that we have not compared the theoretical anisotropies to the 
experimental ones. For small values
of momentum transfer the calculated values are even in qualitative disagreement
with the experimental results, although for large momentum transfer the
qualititative agreement is restored. This result is not surprising, however,
because, as Berggren et al. have argued~\cite{exp-dcp} in their detailed
study, the proper description of the Compton anisotropy mandates 
a good description of the long-range tails of the crystal orbitals. To be
able to do so with the Gaussian-type of basis functions used here, one 
will---unlike the present study---have to include basis functions with quite 
diffuse exponents.  
\section{CONCLUSIONS}
\label{conclusion}
 In conclusion, an {\em ab initio} Hartree-Fock approach for an infinite
insulating  crystal which yields orbitals in a localized representation 
has been discussed in detail. It was applied to compute the total energies
per unit cell, x-ray structure factors and directional Compton profiles
of two halides of lithium, LiF and LiCl. The close agreement between
the results obtained using the present approach, and the ones obtained using
the conventional Bloch orbital based HF approach, demonstrates that the
two approaches are entirely equivalent. The advantage of our approach is
that by considering local perturbations to the Hartree-Fock reference state
by conventional quantum-chemical methods, one can go beyond the mean-field
level and study the influence of electron correlations on an infinite solid 
in an entirely {\em ab initio} manner. Presently projects along
this direction are at progress in our group, and in a future publication
we will study the influence of electron correlations on the ground state
of a solid.
\acknowledgements
One of us (A.S.) gratefully acknowledges useful discussions with Prof. Roberto
Dovesi, and his help regarding the use of the CRYSTAL program.
\appendix
\section{Integral Evaluation}
\label{app1}
 In this section we discuss the calculation of various terms in the Fock
matrix. Since the kinetic-energy matrix elements $T_{pq}=<p|T|q>$ and
the overlap-matrix elements $S_{pq}=<p|q>$ have simple mathematical
expressions and are essentially unchanged from molecular calculations, we
will not discuss them in detail. However, we will consider the 
evaluation of the rest of the contributions to the Fock matrix at some length.
\subsection{Nuclear Attraction Integrals}
\label{app-enuc}
The electron-nucleus attraction term of the Fock matrix contains the 
infinite lattice sums involving the attractive interaction acting on the
electrons of the reference cell due to the infinite number of nuclei
in the solid. When treated individually, this term is divergent. However, 
when combined with the Coulombic part of the electron repulsion to be discussed
in the next section, convergence is achieved because the divergences 
inherent in both sums cancel each other owing to the opposite signs.
This fact is a consequence of the charge neutrality of the unit cell and
is used in the Ewald-summation technique~\cite{ewald} to make the individual 
contibutions also 
convergent by subtracting, from the corresponding potential a shadow
potential emerging from a ficitious homogeneous charge distribution of 
opposite sign. In addition, in the Ewald method, one splits the lattice
potential into a short-range part whose contribution is rapidly convergent
in the ${\bf r}$-space and a long-range part which converges fast
in ${\bf k}$-space. Therefore, in the Ewald-summation technique one
replaces the electron-nucleus interaction potential due to a lattice
composed of nuclei of charge $Z$, by the effective potential~\cite{ewald} 
\begin{equation}
U^{Ew}({\bf r })= -Z\left\{ \sum_{{\bf R}_{i}} \frac{\mbox{erfc}
(\sqrt{\lambda}|{\bf r }-{\bf R}_{i}|) }{|{\bf r }-{\bf R}_{i}|}
+ \frac{4 \pi }{\omega} \sum_{{\bf K}_{i} \neq 0} 
\frac{\mbox{exp}(-\frac{{\bf K}_{i}^{2}} 
{4\lambda}+ i{\bf K}_{i}\cdot {\bf r})}{{\bf K}_{i}^{2}} - \frac{\pi }{\omega}
\frac{1}{\lambda} \right\}
\mbox{,}
\label{eq-ewaldp} 
\end{equation}
where ${\bf R}_{i}$ represents the positions of the nuclei on the lattice,
${\bf K}_{i}$ are the vectors of the reciprocal lattice, $\omega$ is the
volume of the unit cell, $\lambda$ is a convergence parameter to be discussed
later and erfc represents the complement of the error function. Matrix 
elements of the Ewald potential of Eq.(\ref{eq-ewaldp}) with respect to 
primitive $s$-type basis functions were derived by 
Stoll~\cite{ew-stoll} to be 
\begin{equation}
U^{Ew}_{pq}({\bf R}_{p},{\bf R}_{q}) = 
<p({\bf R}_{p})|U^{Ew}|q({\bf R}_{q})> = \hat{U}_{p  q }
 S_{p q } \; \mbox{.}  
\label{eq-upq} 
\end{equation} 
Above $p$ and $q$ label the primitive basis functions, ${\bf R}_{p}$ and
${\bf R}_{q}$ represent the positions of the unit cells in which they are 
located
and $S_{p q}$ represents the overlap matrix element between the two
primitives given by
\begin{equation} 
 S_{p q} =  \frac{2^{3/2}(\eta_{p}\eta_{q})^{3/4}}
{(\eta_{p} + \eta_{q})^{3/2}}\mbox{exp}(-A_{pq}
({\bf r}_{p}+{\bf R}_{p} - {\bf r}_{q}-{\bf R}_{q})^{2}) 
\mbox{.}
\label{eq-spq} 
\end{equation}
The vectors ${\bf r}_{p}$ and ${\bf r}_{q}$ above specify the centers of
 the two basis
functions relative to the origin of the unit cell,
 $\eta_{p}$ and $\eta_{q}$
represent the exponents of the two Gaussians, 
$A_{pq}= \frac{\eta_{p}\eta_{q}}{\eta_{p}+\eta_{q}}$ and 
\begin{equation} 
 \hat{U}_{p q } = -Z W(C_{pq}, {\bf r}_{p,q}) \; \mbox{,}
\label{eq-hupq}
\end{equation} 
with $C_{pq} = 
\eta_{p} + \eta_{q}$, ${\bf r}_{p,q} = \{\eta_{p} ({\bf r}_{p}+{\bf R}_{p})+ 
\eta_{q} ({\bf r}_{q}+{\bf R}_{q})\} C_{pq}^{-1}$ and 

\begin{eqnarray}
W(\alpha, {\bf r}) & = & \sum_{{\bf R}_{i}} \frac{\mbox{erfc}
(\sqrt{\epsilon}|{\bf r }-{\bf R}_{i}|) - \mbox{erfc}
(\sqrt{\alpha}|{\bf r }-{\bf R}_{i}|)}{|{\bf r }-{\bf R}_{i}|}
+\frac{4 \pi}{\omega} \sum_{{\bf K}_{i} \neq 0} 
\frac{\mbox{exp}(-\frac{{\bf K}_{i}^{2}} 
{4\epsilon}+ i{\bf K}_{i}\cdot {\bf r})}{{\bf K}_{i}^{2}} 
 \nonumber \\
             &   & 
 - \frac{\pi}{\omega}\left(\frac{1}{\epsilon} - \frac{1}{\alpha}\right)
\; \mbox{.}
\label{eq-w}      
\end{eqnarray} 
where the parameter $\epsilon$ takes over the role of the convergence
parameter $\lambda$ of Eq.(\ref{eq-ewaldp}). % by $\frac{1}{\lambda}=
%\frac{1}{\epsilon} - \frac{1}{\alpha}$.
The remaining quantities are the
the same as those in Eq.(\ref{eq-ewaldp}).
It is clear that the function
$W(\alpha, {\bf r})$ involves lattice sums both in the direct space and the
reciprocal space. Although the final value of the function
will be independent of the choice of the convergence
parameter $\epsilon$, both these sums can be made to converge optimally
by making a judicious choice of it. Large values of $\epsilon$ lead to
faster convergence in the real space but to slower one in the reciprocal
space and with smaller values of $\epsilon$ the situation is just the 
opposite. Therefore, for optimal performance, the choice of $\epsilon$
is made dependent on the value of $\alpha$. In the present work we make
the choice so that
if $\alpha > 
\frac{\pi}{\omega^{2/3}}$, $\epsilon = \frac{\pi}{\omega^{2/3}}$ and
if  $\alpha \leq 
\frac{\pi}{\omega^{2/3}}$, $\epsilon = \alpha$. In the former case the
sum is both, in the real and the reciprocal space while in the latter case
the sum is entirely in the reciprocal space. Although we have written an
efficient computer code
to evaluate the function $W(\alpha, {\bf r})$, it remains the most computer
intensive part of our program.

The computational effort involved in the computation of these
integrals can be reduced by utilizing the translational symmetry. One can 
verify that as
a consequence of translation symmetry
\begin{equation}
U^{Ew}_{pq}({\bf R}_{p},{\bf R}_{q}) = U^{Ew}_{pq}({\bf t}_{pq},o) 
= U^{Ew}_{pq}({\bf t}_{pq}) \;
\mbox{,}
\label{eq-usym} 
\end{equation} 
where ${\bf t}_{pq}={\bf R}_{p} - {\bf R}_{q}$ is also a vector of the direct
lattice, $o$ represents the reference unit cell and the last term is a
compact notation for the second term. Since the number of unique
${\bf t}_{pq}$ vectors is much smaller than the number of pairs 
$({\bf R}_{p},{\bf R}_{q})$, the use of Eq.(\ref{eq-usym}) reduces the
computational effort considerably. To further reduce the computational
effort we also use the interchange symmetry
\begin{equation}
U^{Ew}_{pq}({\bf t}_{pq}) = U^{Ew}_{qp}(-{\bf t}_{pq})
\; \mbox{.}
\label{eq-usym2}
\end{equation}  
Additional savings are achieved if one
realizes that matrix elements $U^{Ew}_{pq}({\bf t}_{pq})$ become smaller 
as larger the distance $|{\bf t}_{pq}|$ between the interacting charge
distributions becomes. As is clear from Eq.(\ref{eq-upq}), a good estimate of
the magnitude of an integral is the overlap element 
$S_{p  q }$~\cite{ahlrichs}. Therefore, we compute only those
integrals whose overlap elements $S_{p  q }$ are larger than some 
threshold $t_{n}$. In the present calculations we chose $t_{n}=1.0 \times 10^{-7}$.

\subsection{Electronic Coulomb Integrals}
\label{app-clm}

To calculate the Coulomb contribution to the Fock matrix, one needs
to evaluate the two-electron integrals with infinite lattice sum 
\begin{equation}
J_{pq;rs}({\bf R}_{p},{\bf R}_{q},{\bf R}_{r},{\bf R}_{s}) = 
\sum_{k} <p({\bf R}_{p}) \:r({\bf R}_{r}+{\bf R}_{k}) |\frac{1}{r_{12}}
|q({\bf R}_{q})  \: s({\bf R}_{s}+{\bf R}_{k}) >
\mbox{,}
\label{eq-jmat}
\end{equation}
where $p, q, r$ and $s$ represent the primitive basis functions and 
${\bf R}_{p}, {\bf R}_{q}, {\bf R}_{r}$ and ${\bf R}_{s}$ represent the 
unit cells in which they are centered. This integral, treated on its own is 
divergent, as discussed in the previous section. However, using the
Ewald-summation technique, one can make this series conditionally convergent
with the implicit assumption that its divergence will cancel the corresponding
divergence of the electron nucleus interaction. Since the details of
the Ewald-summation technique for the Coulomb part of electron repulsion
 are essentially identical to the case of electron-nucleus interaction,
 we will just state the final results~\cite{ew-stoll}
\begin{equation}
\tilde{J}_{pq;rs}({\bf R}_{p},{\bf R}_{q},{\bf R}_{r},{\bf R}_{s}) = 
S_{p  q } S_{r  s } W(B_{rs}^{pq}, {\bf r}^{r,s}_{p,q})
\label{eq-jmate}
\end{equation}
where
$$(B^{pq}_{rs})^{-1} = 
   (\eta_{p} + \eta_{q})^{-1} + (\eta_{r} + \eta_{s})^{-1} \; \mbox{,} $$
and 
 $$    {\bf r}^{r,s}_{p,q}     =  {\bf r}_{r,s}- {\bf r}_{p,q} \; \mbox{.}$$

All the notations used in the equations above were defined in the previous
section. The expression $\tilde{J}_{pq;rs}$ used in Eq.(\ref{eq-jmate}),
as against $J_{pq;rs}$ of Eq.(\ref{eq-jmat}), is meant to remind us
that the matrix elements stated in  Eq.(\ref{eq-jmate}) are those of 
the two-electron  Ewald potential and not those of the ordinary 
Coulomb potential.
 
Like in the case of electron-nucleus attraction, one can utilize 
the translational symmetry for the present case to reduce the computational
effort significantly. The corresponding relations in the present case are
\begin{equation}
\tilde{J}_{pq;rs}({\bf R}_{p},{\bf R}_{q},{\bf R}_{r},{\bf R}_{s}) = 
\tilde{J}_{pq;rs}({\bf t}_{pq},o,{\bf t}_{rs},o) = 
\tilde{J}_{pq;rs}({\bf t}_{pq},{\bf t}_{rs})\mbox{,}
\label{eq-jsym}
\end{equation}
where as before $o$ represents the reference unit cell, 
${\bf t}_{pq}={\bf R}_{p} - {\bf R}_{q}$, ${\bf t}_{rs}={\bf R}_{r} - {\bf
  R}_{s}$ and the last term in Eq.(\ref{eq-jsym}) is a compact notation
for the second term. Since the number of pairs $({\bf t}_{pq},{\bf t}_{rs})$
is much smaller than the number of quartets 
$({\bf R}_{p},{\bf R}_{q},{\bf R}_{r},{\bf R}_{s})$, use of Eq.(\ref{eq-jsym})
results in considerable savings of computer time and memory. In addition,
we also use the four interchange relations of the form of Eq.(\ref{eq-usym2})
to further reduce the number of nonredundant integrals. Additionally, these
integrals also satisfy the interchange relation
\begin{equation}
\tilde{J}_{pq,rs}({\bf t}_{pq},{\bf t}_{rs}) = 
\tilde{J}_{rs,pq}({\bf t}_{rs},{\bf t}_{pq}) \; \mbox{.}
\label{eq-jsymn} 
\end{equation}
To keep the programming simple, however, at present we do not utilize this
symmetry. In future, we do intend to incorporate this symmetry in the code. 

Similar to the case of electron-nucleus integrals, here also
we use the magnitude of the product $S_{p  q } S_{r  s }$ to 
estimate the
size of the integral to be computed and  proceed with its calculation
only if it is greater than a threshold $t_{c}$, taken to be $1.0\times 10^{-7}$
in this study.
\subsection{Electronic Exchange Integrals}
\label{app-exc}
In order to compute the exchange contribution to the Fock matrix, one has to
compute the following two-electron integrals involving infinite lattice sum
\begin{equation}
K_{pq;rs}({\bf R}_{p},{\bf R}_{q},{\bf R}_{r},{\bf R}_{s}) = 
\sum_{k} <p({\bf R}_{p}) \: s({\bf R}_{s}+{\bf
  R}_{k})|\frac{1}{r_{12}}|r({\bf R}_{r}+{\bf R}_{k}) \: q({\bf R}_{q}) >
\mbox{,}
\label{eq-kmat}
\end{equation}
where the notation is identical to the previous two cases.
 By using the translational symmetry arguments one can show even for the
exchange case that
\begin{equation}
K_{pq;rs}({\bf R}_{p},{\bf R}_{q},{\bf R}_{r},{\bf R}_{s}) = 
K_{pq;rs}({\bf t}_{pq},o,{\bf t}_{rs},o) = K_{pq;rs}({\bf t}_{pq},{\bf t}_{rs})\mbox{,}
\label{eq-ksym}
\end{equation}
 where the last term in Eq.(\ref{eq-ksym}) above is a compact notation
for the second term. As in the previous two cases, the use of translational
symmetry results in considerable savings of computer time and storage.
Explicitly
\begin{equation}
K_{pq;rs}({\bf t}_{pq},{\bf t}_{rs}) =  
\sum_{k} <p({\bf t}_{pq}) \: s({\bf
  R}_{k})|\frac{1}{r_{12}}|r({\bf t}_{rs}+{\bf R}_{k}) \: q(o) >
\mbox{.}
\label{eq-kfin}
\end{equation}
Although Eq.(\ref{eq-kfin}) contains an infinite sum over lattice 
vectors
${\bf R}_{k}$, the contributions of each of the terms decreases rapidly with
the increasing distances $|{\bf t}_{rs}+{\bf R}_{k}-{\bf
  t}_{pq}|$ and $|{\bf R}_{k}|$ between the interacting charge
distributions. A good
estimate of the contribution of the individual terms is provided by the
product of overlap matrix elements between the interacting charge distributions
namely, $S_{pr}=<p({\bf t}_{pq})|r({\bf t}_{rs}+{\bf R}_{k})>$
and $S_{qs}= <q(o)|s({\bf R}_{k})>$~\cite{ahlrichs}.
Therefore, in the computer implementation, we arrange the vectors ${\bf R}_{k}$
so that the corresponding overlaps are in the descending order and the
loop involving the sum over  ${\bf R}_{k}$ in Eq.(\ref{eq-kfin}) is 
terminated once the individual overlap matrix elements or their product are
less than a specified threshold $t_{e}$. 
The computer code for evaluating these integrals is a modified version of the 
program written originally by Ahlrichs~\cite{ahlrichs}. The value of the 
threshold $t_{e}$ used in these calculations was $1.0\times 10^{-7}$.
The exchange integrals also satisfy interchange symmetries similar to those
of Eqs.(\ref{eq-usym2}) and (\ref{eq-jsymn}), which are not used in the
present version of the code for the ease of programming. In future, however,
we plan to use them as well.

As described above, to minimize the need of computer time and storage, we have
made extensive use of translational symmetry. However, the integral
evaluation can be further optimized considerably by making use of point
group symmetry as is done in the CRYSTAL program~\cite{crystalprog}.
Implementation of point group symmetry, as well as the use of CGTOs instead
of lobe-type functions, is planned for future improvements of the present
code.

\begin{figure}
\caption{LiF: $2p_{z}$-type valence Wannier function centered on 
$\mbox{F}^{-}$  (located at origin) plotted along the ${\bf r} = x(0,0,1)$
direction. Nodes near the $(0,0,\pm \frac{a}{2})$ positions are due to its
orthogonalization to the Li$^{+}$ $1s$ orbital located there. All distances
are in atomic units.}
\label{fig-lif2p-001} 
\end{figure}
\begin{figure}
\caption{LiF: $2p_{x}$-type valence Wannier function centered on 
$\mbox{F}^{-}$ (located at origin) plotted along the ${\bf r} = x(1,1,0)$
direction. Nodes near the $(\pm \frac{a}{2},\pm \frac{a}{2},0)$ positions are 
due to its orthogonalization to $2p$ orbitals of $\mbox{F}^{-}$ located 
there. All distances
are in atomic units.}
\label{fig-lif2p-110} 
\end{figure} 
\begin{figure}
\caption{LiCl: $3p_{z}$-type valence Wannier function centered on 
$\mbox{Cl}^{-}$ (located at origin) plotted along the ${\bf r} = x(0,0,1)$
direction. Nodes near the origin are due to its orthogonalization to
the $\mbox{Cl}^{-}$ $2p$-orbitals centered there while those  
 near the $(0,0,\pm \frac{a}{2})$ positions are due to its
orthogonalization to the Li$^{+}$ $1s$ orbital located there. All distances
are in atomic units.}
\label{fig-licl3p-001} 
\end{figure}
\begin{figure}
\caption{LiCl: $3p_{x}$-type valence Wannier function centered on 
$\mbox{Cl}^{-}$ (located at origin) plotted along the ${\bf r} = x(1,1,0)$
direction. Nodes near the origin are due to its orthogonalization to
the $\mbox{Cl}^{-}$ $2p$-orbitals centered there, while the two nodes each 
near the  $(\pm \frac{a}{2},\pm \frac{a}{2},0)$ positions are due to its
orthogonalization to both, the $3s$ and the $3p$ orbitals of $\mbox{Cl}^{-}$  
located there. All distances are in atomic units.}
\label{fig-licl3p-110} 
\end{figure} 
\begin{table}  
 \protect\caption{Exponents and contraction coefficients used in the
basis set for lithium\protect\cite{lih-dovesi}. }
 \protect\begin{center}  
  \begin{tabular}{lrl} \hline     
 \multicolumn{1}{c}{Shell type} 
   & \multicolumn{1}{c}{Exponent}
    & \multicolumn{1}{c}{Contraction Coefficient} \\ \hline
 1s  &     700.0   &  0.001421 \\
     &     220.0   &  0.003973 \\
     &     70.0    &  0.016390 \\
     &     20.0    &  0.089954  \\
     &     5.0     &  0.315646  \\
     &     1.5     &  0.494595 \\ \hline
 2s  &     0.5     &  1.0  \\ \hline
 2p  &     0.6     &  1.0  \\ \hline 
   \end{tabular}                      
   \end{center}  
  \label{tab-basli}    
\end{table}  
\begin{table}  
 \protect\caption{Exponents and contraction coefficients used in the
basis set for fluorine~\protect\cite{huzbas}. }
 \protect\begin{center}  
  \begin{tabular}{lll} \hline     
 \multicolumn{1}{c}{Shell type} 
   & \multicolumn{1}{c}{Exponent}
    & \multicolumn{1}{c}{Contraction Coefficient} \\ \hline
 1s  &  2931.321 & 0.005350 \\
     &  441.9897 & 0.039730 \\
     &  100.7312 & 0.177257 \\
     &  28.14426 & 0.457105 \\ \hline
 2s  &    8.7256 & 1.0      \\ \hline
 3s  & 1.40145   & 1.0 \\ \hline
 4s  & 0.41673   & 1.0 \\ \hline  
 2p  & 10.56917  & 0.126452 \\
     & 2.19471   & 0.478100 \\ \hline
 3p  & 0.47911   & 1.0 \\ \hline 
   \end{tabular}                      
   \end{center}  
  \label{tab-basf}    
\end{table}  
\begin{table}  
 \protect\caption{Exponents and contraction coefficients used in the
basis set for chlorine~\protect\cite{huzbas}. }
 \protect\begin{center}  
  \begin{tabular}{lll} \hline      
 \multicolumn{1}{c}{Shell type} 
   & \multicolumn{1}{c}{Exponent}
    & \multicolumn{1}{c}{Contraction Coefficient} \\ \hline
 1s  & 30008.27 & 0.001471 \\
     & 4495.692 & 0.011324 \\ 
     & 1021.396 & 0.056401 \\
     & 287.6894 & 0.200188 \\
     & 92.26777 & 0.443036 \\
     & 31.76476 & 0.402714 \\ \hline
 2s  & 7.16468  & 1.0 \\ \hline
 3s  & 2.78327  & 1.0 \\ \hline
 4s  & 0.60063  & 1.0 \\ \hline
 5s  & 0.22246  & 1.0 \\ \hline
 2p  & 157.7332 & 0.025920 \\
     & 36.27829 & 0.164799 \\
     & 10.84    & 0.460043 \\
     & 3.49773  & 0.499410 \\ \hline
 3p  & 0.77581  & 1.0   \\ \hline
 4p  & 0.21506  & 1.0 \\ \hline
   \end{tabular}                      
   \end{center}  
  \label{tab-bascl}    
\end{table}  
\begin{table}  
 \protect\caption{Comparison between between total energies obtained using
our approach and those obtained using
 CRYSTAL\protect\cite{crystalprog} for lithium fluoride for different
values of lattice constants. The ${\cal N}$ region included up 
to third-nearest neighbor unit cells. Lattice constants are in
units of \AA, and energies are in atomic units.}
% \protect\begin{center}  
  \begin{tabular}{|llr|} \hline     
 Lattice Constant & \multicolumn{2}{c|}{Total Energy}  \\  \hline 
              & This work   & CRYSTAL \\ \hline 
   3.8        &-106.8985    & -106.8980 \\
   3.9        &-106.8939    & -106.8935       \\
   3.99       &-106.8877    & -106.8873    \\  
   4.1        &-106.8780    & -106.8774   \\
   4.2        &-106.8677    & -106.8670    \\
\hline  
   \end{tabular}                      
%   \end{center}  
  \label{tab-enlif}    
\end{table}  
\begin{table}  
 \protect\caption{Comparison between between total energies obtained using
our approach and those obtained using
 CRYSTAL\protect\cite{crystalprog} for lithium chloride for different
values of lattice constants. The ${\cal N}$ region included up 
to third-nearest neighbor unit cells. Lattice constants are in
units of \AA, and energies are in atomic units.}
% \protect\begin{center}  
  \begin{tabular}{|llr|} \hline     
 Lattice Constant & \multicolumn{2}{c|}{Total Energy}  \\  \hline 
              & This work   & CRYSTAL \\ \hline 
   4.9        &-466.5062    & -466.5065    \\
   5.0        &-466.5078    & -466.5082       \\
   5.07       &-466.5080    & -466.5085    \\  
   5.2        &-466.5065    & -466.5071   \\
   5.3        &-466.5041    & -466.5047   \\
\hline  
   \end{tabular}                      
%   \end{center}  
  \label{tab-enlicl}    
\end{table}  
\begin{table}  
 \protect\caption{Calculated and experimental values of x-ray structure
factors for LiF in electrons per unit cell. The experimental structure factors
are taken from reference\protect\cite{exp-xraylif}. The Debye-Waller 
corrections were removed\protect\cite{exp-xraylifeuw}. The reciprocal-lattice 
vectors are defined with respect to
the conventional cubic unit cell and not the primitive cell. They are labeled
by integers $h$, $k$ and $l$.}
% \protect\begin{center}  
  \begin{tabular}{cccc} \hline     
 $hkl$ & Experimental & This work & CRYSTAL  \\ \hline 
 111   & 4.84         & 5.04      & 5.04   \\
 200   & 7.74         & 7.78      & 7.78   \\ 
 220   & 5.71         & 5.68      & 5.68   \\
 311   & 2.37         & 2.32      & 2.32    \\
 222   & 4.61         & 4.52      & 4.52    \\
 400   & 3.99         & 3.84      & 3.84    \\
 331   & 1.65         & 1.60      & 1.60    \\
 420   & 3.46         & 3.35      & 3.35    \\
 422   & 3.07         & 2.99      & 2.99    \\
 511   & 1.38         & 1.34      & 1.33    \\
 333   & 1.38         & 1.33      & 1.33    \\
 440   & 2.58         & 2.52      & 2.52     \\
 531   & 1.28         & 1.22      & 1.22     \\
 600   & 2.41         & 2.36      & 2.35     \\
 442   & 2.41         & 2.35      & 2.35     \\
 620   & 2.24         & 2.22      & 2.22      \\  
   \end{tabular}                      
%   \end{center}  
  \label{tab-lifxfac}    
\end{table}  
\begin{table}  
 \protect\caption{Calculated and experimental values of x-ray structure
factors for LiCl in electrons per unit cell. Second and third columns report 
the theoretical values obtained by the specified method, without including 
the Debye-Waller corrections.
Next column reports the theoretical values after including the Debye-Waller
factors of $B_{\mbox{Li}}=0.93 \; \mbox{\AA}^{2}$ and $B_{\mbox{Cl}}=0.41 \;
\mbox{\AA}^{2}$   corresponding to a temperature 
$T= 78 \; K $\protect\cite{exp-xraylicl}. Last column
reports  experimental values of x-ray structure factors mesured at 
$T= 78 \; K$\protect\cite{exp-xraylicl}.  
The reciprocal-lattice vectors are defined with respect to
the conventional cubic unit cell and not the primitive cell. They are labeled
by integers $h$, $k$ and $l$.}
% \protect\begin{center}  
  \begin{tabular}{ccccc} \hline     
\multicolumn{1}{l}{} & \multicolumn{2}{c}{Uncorrected} & 
      \multicolumn{1}{c}{Debye-Waller Corrected} &  \multicolumn{1}{c}{} \\ 
 $hkl$ & This Work & CRYSTAL & This work & Experimental  \\ \hline 
 111   & 11.28     & 11.28   & 11.18     & 10.91         \\
 200   & 13.96     & 13.96   & 13.70     & 13.77         \\ 
 220   & 11.46     & 11.46   & 11.04     & 11.03     \\
 311   & 7.55      & 7.55    & 7.30      & 7.44    \\
 222   & 10.20     & 10.20   & 9.64      & 9.76     \\
 400   & 9.43      & 9.44    & 8.76      & 8.95     \\
 331   & 6.61      & 6.62    & 6.23      & 6.24     \\
 420   & 8.87      & 8.88    & 8.09      & 8.15     \\
 422   & 8.43      & 8.43    & 7.55      & 7.60     \\
 511   & 6.15      & 6.16    & 5.64      & 5.61  \\
 333   & 6.15      & 6.16    & 5.64      & 5.61  \\
 440   & 7.73      & 7.74    & 6.69      & 6.70   \\
 531   & 5.81      & 5.81    & 5.17      & 5.30   \\
 600   & 7.44      & 7.44    & 6.32      & 6.53   \\
 442   & 7.43      & 7.44    & 6.32      & 6.53   \\
 620   & 7.16      & 7.17    & 5.99      & 5.95   \\  
   \end{tabular}                      
%   \end{center}  
  \label{tab-liclxfac}    
\end{table}  
\begin{table}
% \protect\begin{center}  
 \protect\caption{Theoretical HF directional Compton profiles for LiF of 
this work ($J_{TW}$) compared to those of CRYSTAL ($J_{CR}$). The
directionally averaged Compton profiles of both the approaches 
($<J_{TW}>$ and $<J_{CR}>$) are also compared to the experimental isotropic
Compton profiles ($J_{exp})$\protect\cite{exp-cp}. 
The Compton profiles and momentum transfer $q$ are in atomic units. The
column headings [$h$$k$$l$]  refer to the direction of momentum transfer in 
the crystal. All the profiles are normalized to 5.865 electrons in the
interval $q=0-7$ a.u.}
  \begin{tabular}{lrlrlrlrlc} \hline
 \multicolumn{1}{l}{} & \multicolumn{2}{c}{[100]} 
    & \multicolumn{2}{c}{[110]} & \multicolumn{2}{c}{[111]} &
 \multicolumn{2}{c}{average} & \multicolumn{1}{c}{ }\\     
 $q$ & $J_{TW}$ 
             & $J_{CR}$  
                     & $J_{TW}$ 
                             & $J_{CR}$                      
                                     & $J_{TW}$ 
                                             & $J_{CR}$  
                                                     & $<J_{TW}>$ 
                                                             & $<J_{CR}>$  
                                                                     & $J_{exp}$
     \\ \hline 
 0.0 & 3.759 & 3.762 & 3.762 & 3.760 & 3.777 & 3.774 & 3.766 & 3.764 & 3.832 \\
 0.1 & 3.741 & 3.743 & 3.749 & 3.746 & 3.762 & 3.759 & 3.751 & 3.749 & 3.814 \\
 0.2 & 3.689 & 3.691 & 3.707 & 3.705 & 3.718 & 3.715 & 3.706 & 3.705 & 3.765 \\
 0.3 & 3.609 & 3.609 & 3.638 & 3.636 & 3.644 & 3.641 & 3.633 & 3.632 & 3.684 \\
 0.4 & 3.504 & 3.504 & 3.541 & 3.540 & 3.542 & 3.540 & 3.532 & 3.531 & 3.574 \\
 0.5 & 3.382 & 3.382 & 3.416 & 3.415 & 3.413 & 3.411 & 3.407 & 3.406 & 3.434 \\
 0.6 & 3.245 & 3.245 & 3.266 & 3.266 & 3.258 & 3.257 & 3.259 & 3.258 & 3.271 \\
 0.7 & 3.095 & 3.094 & 3.094 & 3.093 & 3.081 & 3.081 & 3.090 & 3.090 & 3.089 \\
 0.8 & 2.929 & 2.928 & 2.901 & 2.901 & 2.886 & 2.887 & 2.903 & 2.903 & 2.886 \\
 0.9 & 2.745 & 2.745 & 2.692 & 2.692 & 2.677 & 2.678 & 2.700 & 2.700 & 2.662 \\
 1.0 & 2.541 & 2.541 & 2.472 & 2.473 & 2.458 & 2.460 & 2.484 & 2.485 & 2.426 \\
 1.2 & 2.078 & 2.077 & 2.022 & 2.025 & 2.020 & 2.022 & 2.035 & 2.036 & 1.948 \\
 1.4 & 1.608 & 1.606 & 1.606 & 1.607 & 1.616 & 1.618 & 1.610 & 1.610 & 1.530 \\
 1.6 & 1.224 & 1.224 & 1.261 & 1.260 & 1.275 & 1.276 & 1.257 & 1.257 & 1.202 \\
 1.8 & 0.957 & 0.956 & 0.994 & 0.995 & 1.003 & 1.003 & 0.988 & 0.988 & 0.955 \\
 2.0 & 0.772 & 0.771 & 0.797 & 0.797 & 0.795 & 0.795 & 0.790 & 0.791 & 0.778 \\
 3.0 & 0.339 & 0.338 & 0.324 & 0.325 & 0.329 & 0.329 & 0.329 & 0.329 & 0.336 \\
 3.5 & 0.236 & 0.236 & 0.244 & 0.244 & 0.241 & 0.240 & 0.241 & 0.241 & 0.243 \\
 4.0 & 0.179 & 0.179 & 0.181 & 0.181 & 0.182 & 0.182 & 0.181 & 0.181 & 0.188 \\
 5.0 & 0.112 & 0.113 & 0.113 & 0.113 & 0.112 & 0.112 & 0.113 & 0.113 & 0.115 \\
 6.0 & 0.074 & 0.074 & 0.074 & 0.074 & 0.074 & 0.074 & 0.074 & 0.074 & 0.077 \\
 7.0 & 0.050 & 0.050 & 0.050 & 0.050 & 0.050 & 0.050 & 0.050 & 0.050 & 0.051 \\
\hline 
   \end{tabular}                      
  \label{tab-lifcp}    
\end{table}  
\begin{table}
% \protect\begin{center}  
 \protect\caption{Theoretical HF directional Compton profiles for LiCl of 
this work ($J_{TW}$) compared to those of CRYSTAL ($J_{CR}$). The
directionally averaged Compton profiles of both the approaches 
($<J_{TW}>$ and $<J_{CR}>$) are also compared to the experimental isotropic
Compton profiles ($J_{exp})$\protect\cite{exp-cp}. 
The Compton profiles and momentum transfer $q$ are in atomic units. The
column headings [$h$$k$$l$]  refer to the direction of momentum transfer in 
the crystal. All the profiles are normalized to 9.365 electrons in the
interval $q=0-7$ a.u.}
  \begin{tabular}{lrlrlrlrlc} \hline
 \multicolumn{1}{l}{} & \multicolumn{2}{c}{[100]} 
    & \multicolumn{2}{c}{[110]} & \multicolumn{2}{c}{[111]} &
 \multicolumn{2}{c}{average} & \multicolumn{1}{c}{ }\\     
 $q$ & $J_{TW}$ 
             & $J_{CR}$  
                     & $J_{TW}$ 
                             & $J_{CR}$                      
                                     & $J_{TW}$ 
                                             & $J_{CR}$  
                                                     & $<J_{TW}>$ 
                                                             & $<J_{CR}>$  
                                                                     & $J_{exp}$
     \\ \hline 
 0.0 & 6.190 & 6.209 & 6.207 & 6.198 & 6.217 & 6.204 & 6.206 & 6.202 & 6.282 \\
 0.1 & 6.152 & 6.169 & 6.173 & 6.166 & 6.181 & 6.169 & 6.171 & 6.168 & 6.228 \\
 0.2 & 6.041 & 6.051 & 6.066 & 6.065 & 6.073 & 6.064 & 6.062 & 6.062 & 6.100 \\
 0.3 & 5.861 & 5.864 & 5.881 & 5.883 & 5.892 & 5.887 & 5.879 & 5.880 & 5.896 \\
 0.4 & 5.613 & 5.607 & 5.617 & 5.619 & 5.634 & 5.633 & 5.622 & 5.620 & 5.613 \\
 0.5 & 5.297 & 5.286 & 5.289 & 5.286 & 5.302 & 5.305 & 5.295 & 5.292 & 5.262 \\
 0.6 & 4.919 & 4.910 & 4.903 & 4.900 & 4.903 & 4.908 & 4.907 & 4.904 & 4.857 \\
 0.7 & 4.488 & 4.486 & 4.472 & 4.473 & 4.452 & 4.457 & 4.469 & 4.471 & 4.416 \\
 0.8 & 4.023 & 4.028 & 4.007 & 4.014 & 3.972 & 3.978 & 4.000 & 4.006 & 3.958 \\
 0.9 & 3.544 & 3.552 & 3.528 & 3.539 & 3.493 & 3.500 & 3.521 & 3.530 & 3.512 \\
 1.0 & 3.081 & 3.086 & 3.065 & 3.075 & 3.046 & 3.053 & 3.063 & 3.071 & 3.100 \\
 1.2 & 2.309 & 2.308 & 2.304 & 2.305 & 2.326 & 2.328 & 2.312 & 2.313 & 2.403 \\
 1.4 & 1.817 & 1.817 & 1.827 & 1.825 & 1.850 & 1.848 & 1.832 & 1.830 & 1.897 \\
 1.6 & 1.534 & 1.532 & 1.549 & 1.545 & 1.548 & 1.546 & 1.545 & 1.542 & 1.570 \\
 1.8 & 1.350 & 1.347 & 1.361 & 1.358 & 1.349 & 1.347 & 1.355 & 1.352 & 1.374 \\
 2.0 & 1.213 & 1.212 & 1.210 & 1.211 & 1.206 & 1.204 & 1.210 & 1.209 & 1.227 \\
 3.0 & 0.778 & 0.777 & 0.775 & 0.777 & 0.776 & 0.776 & 0.776 & 0.776 & 0.770 \\
 3.5 & 0.630 & 0.629 & 0.632 & 0.630 & 0.631 & 0.631 & 0.631 & 0.630 & 0.608 \\
 4.0 & 0.512 & 0.512 & 0.510 & 0.511 & 0.511 & 0.511 & 0.511 & 0.511 & 0.487 \\
 5.0 & 0.334 & 0.333 & 0.333 & 0.334 & 0.334 & 0.334 & 0.334 & 0.334 & 0.322 \\
 6.0 & 0.224 & 0.224 & 0.224 & 0.225 & 0.224 & 0.225 & 0.224 & 0.224 & 0.213 \\
 7.0 & 0.158 & 0.158 & 0.157 & 0.158 & 0.157 & 0.158 & 0.157 & 0.158 & 0.151 \\
\hline 
   \end{tabular}                      
  \label{tab-liclcp}    
\end{table}  

\begin{references}
\bibitem{au}{We use the conventional atomic units in which the unit of
length is Bohr and the unit of energy is Hartree.}
\bibitem{ref-bas}{The basis functions with $n_{x}+n_{y}+n_{z}=2$ and higher
cannot be characterized as pure spherical harmonics and contain contributions
from lower angular momenta as well. For example, $n_{x}+n_{y}+n_{z}=2$
functions also contain $s$-type contributions besides the contribution
from $d$-type functions. However, in the quantum-chemistry community,
the value of $n_{x}+n_{y}+n_{z}$ is colloquially referred to as the
angular momentum of the basis function.}
\bibitem{denmat}{Usually in quantum-chemistry literature the 
contribution of the two spin projections
is absorbed in the density matrix and it is defined with a factor
of 2 multiplying the right hand side of Eq.(\ref{eq-denmat}). However,
we have chosen to absorb the contributions of the spin sum in various
terms of the Fock matrix and total energy, instead.}
\end{references}
\end{document}